 \documentclass[11pt,a4paper]{article} 
 \usepackage[numbers,sort&compress]{natbib}
 \usepackage{amsmath}
\usepackage{amssymb}
\usepackage{authblk}
\usepackage{graphicx}
\usepackage{amsmath}
\usepackage{comment}
\usepackage{braket}
\usepackage{caption}
\usepackage[toc,page]{appendix}

\title{Quantum teleportation with independent sources over an optical fibre network}

\author[1,2,3]{Qi-Chao Sun}
\author[1,2]{Ya-Li Mao}
\author[4]{Sijing Chen}
\author[5]{Wei Zhang}
\author[1,2]{Yang-Fan Jiang}
\author[6]{Yanbao Zhang}
\author[4]{Weijun Zhang}
\author[7]{Shigehito Miki}
\author[7]{Taro Yamashita}
\author[7]{Hirotaka Terai}
\author[1,2]{Xiao Jiang}
\author[1,2]{Teng-Yun Chen}
\author[4]{Lixing You}
\author[3]{Xianfeng Chen}
\author[4]{Zhen Wang}
\author[1,2]{Jingyun Fan}
\author[1,2]{Qiang Zhang}
\author[1,2]{Jian-Wei Pan}

\affil[1]{  National Laboratory for Physical Sciences at Microscale and Department of Modern Physics, Shanghai Branch, University of Science and Technology of China, Hefei, Anhui 230026, China}
\affil[2]{CAS Center for Excellence and Synergetic Innovation Center in Quantum Information and Quantum Physics, Shanghai Branch, University of Science and Technology of China, Hefei, Anhui 230026, China}
\affil[3]{ Department of Physics and Astronomy, Shanghai Jiao Tong University, Shanghai, 200240, China}
\affil[4]{ State Key Laboratory of Functional Materials for Informatics, Shanghai Institute of Microsystem and Information Technology, Chinese Academy of Sciences, Shanghai 200050, China}
 \affil[5]{ Tsinghua National Laboratory for Information Science and Technology, Department of Electronic Engineering, Tsinghua University, Beijing 100084, China}
 \affil[6]{ Institute for Quantum Computing and Department of Physics and Astronomy, University of Waterloo, Waterloo, Ontario, N2L 3G1 Canada}
\affil[7]{Advanced ICT Research Institute, National Institute of Information and Communications Technology, 588-2, Iwaoka, Nishi-ku, Kobe, Hyogo 651-2492, Japan}

\date{\today}

\begin{document}

\maketitle

Quantum teleportation~\cite{BB:Teleportation} faithfully transfers a quantum state between distant nodes in a network, enabling revolutionary information processing applications\cite{nielsen2010quantum, Cirac:QuantumNetwork, Kimble:QuantumInternet}. Here we report teleporting quantum states over a 30 km optical fibre network with the input single photon state and the EPR state prepared independently. By buffering photons in 10 km coiled optical fibre, we perform Bell state measurement (BSM) after entanglement distribution. With active feed-forward operation, the average quantum state fidelity and quantum process fidelity are measured to be $85(3)\%$ and $77(3)\%$, exceeding classical limits of 2/3 and 1/2, respectively. The statistical hypothesis test shows that the probability of a classical process to predict an average state fidelity no less than the one observed in our experiment is less than  $2.4\times10^{-14}$, confirming the quantum nature of our quantum teleportation experiment. Our experiment marks a critical step towards the realization of quantum internet in the future.

Quantum entanglement is at the heart of quantum mechanics. Bennett et al.~\cite{BB:Teleportation} found that quantum entanglement
is the key to realize the dream of teleportation. Quantum teleportation faithfully transfers the quantum state of a physical system
instead of the system itself between distant nodes. This underlies the proposals of distributed quantum computing~\cite{barz2012demonstration, nielsen2010quantum}
and quantum communication network~\cite{Cirac:QuantumNetwork, Kimble:QuantumInternet}.
\begin{figure}
\centering
\includegraphics[width=0.6\textwidth]{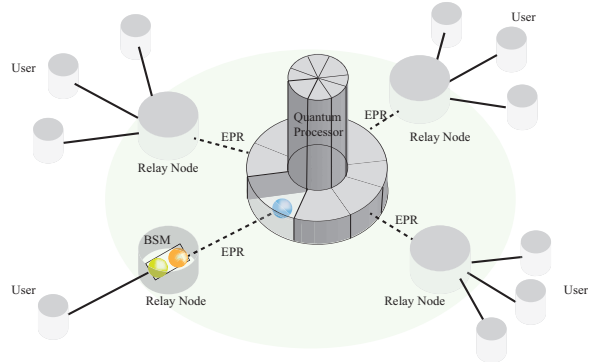}
\caption{\textbf{Schematic of a notional quantum network.} }
\label{fig:network}
\end{figure}

Fig.~\ref{fig:network} depicts a future quantum network. The central node hosts a quantum processor. It shares entanglement with many relay nodes, constituting a star topology structure. The end user accesses the central quantum processor by teleporting the quantum state to the central node via the nearby relay node. The relay nodes perform entanglement distribution and Bell state measurement, and feed-forward the measurement outcomes to the central processor. The structure may be replicated to form a larger network. The quantum network with distant nodes demands that each node has an independent quantum source. Therefore, a critical step in the road map towards realizing a quantum network in the real world is to simultaneously realize independent quantum sources, prior entanglement distribution~ \cite{Gisin:Prior} and active feed-forward operation~\cite{Ma:143Tele} in a single field test of quantum teleportation. This, however, remains a challenge, after many elegant laboratorial demonstrations ~\cite{bouwmeester1997experimental,boschi1998experimental,furusawa1998unconditional,nielsen1998complete, Marcikic:Tele, barrett2004deterministic, riebe2004deterministic, sherson2006quantum, zhang2006experimental, stevenson2013quantum, bussieres2014quantum, hanson2014teleportation, wang2015quantum, takesue2015quantum} and remarkable field tests of quantum teleportation ~\cite{ursin2004communications, Gisin:Prior, Yin:97Tele, Ma:143Tele}. Synchronization of independent quantum sources was notoriously difficult. Previous attempts to synchronize independent quantum sources were within the laboratory~\cite{Tao:Synch, Rainer:Synch, Halder:TimeMeasurement, patel2010two}. Their practicability for field applications is unjustified. To overcome this challenge, we create quantum states with temporal coherence of 110 ps. This allows us to use off-the-shelf instruments to synchronize independent quantum sources with ease. In addition, we buffer photons in coiled optical fibre at both relay node and central node after entanglement distribution~\cite{Gisin:Prior}. This allows us to perform the BSM after photons arrive at the central node and to forward the BSM outcomes in real time to the central node for proper unitary rotation~\cite{Ma:143Tele}. Our quantum teleportation experiment over a realistic network with independent quantum sources, prior entanglement distribution, and active feed-forward operation marks a critical step in the roadmap towards realizing global quantum network.

The network is deployed in the city of Hefei, China. As shown in Fig.~\ref{fig:setup}~a, Charlie (N~$31^{\circ}51^{'}$ $5.42^{''}$, E~$117^{\circ}11^{'}55.82^{''}$) is the relay, Alice (N~$31^{\circ}50^{'}7.50^{''}$, E~$117^{\circ}15^{'}50.56^{''}$) is the user and Bob (N~$31^{\circ}50^{'}11.42^{''}$, E~$117^{\circ}7^{'}54.37^{''}$) is the central processor. Alice (Bob) is connected to Charlie with a piece of 15.7 km (14.7 km) single mode optical fibre with propagation loss of 5 dB (6 dB).

\begin{figure}
\centering
\includegraphics[width=0.9\textwidth]{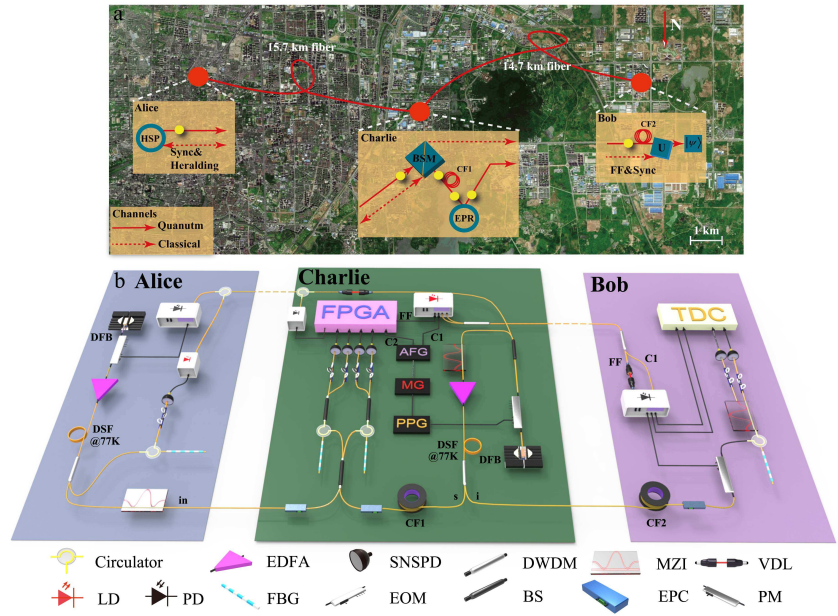}
\caption{\textbf{Quantum teleportation in Hefei optical fibre network.} \textbf{a}, Bird-view of the experiment. Alice prepares quantum state, $|\psi\rangle_{in}$, on a heralded single photon (HSP) and sends it to Charlie, who prior shares an EPR pair with Bob. Each photon of the EPR pair is stored in 15 km coiled fibre (CF). Charlie implements  BSM on his photon and  the received HSP. Then he sends  the feed-forward signal to Bob, who performs  unitary operation (U) and state analysis. The quantum signals are transmitted in optical fibre denoted by solid line, and the classical signals in other optical fibre denoted by dash line.
}
\label{fig:setup}
\end{figure}
\begin{figure}\ContinuedFloat
\centering
\caption*{\textbf{b}, experimental setup. To synchronize the HSP source and EPR source, Charlie sends a portion of each of his laser pulses to Alice through the an optical fibre channel, and Alice detects them by using a 45~GHz photo-detector (PD) to generate driven pulse for electro-optical modulator (EOM). A variable delay line (VDL) is used to adjust the time delay of the laser pulse. The heralding signal is converted into laser pulse with laser diode (LD) and transmitted to Charlie through classical channel. Two optical circulator (Cir) are used to achieve the bi-direction signal transmission. Charlie uses a PD to detect the heralding signal and performs  BSM on the HSP and his photon. The feed-forward signal (FF) and 10~MHz clock (C1, 200 MHz clock is denoted by C2)  are carried by laser pulses of wavelength 1550.92~nm and 1550.12~nm, which are launched into classical channel by a dense wavelength division multiplexing (DWDM) filter and sent to Bob. After separated by another DWDM, the feed-forward signal and 10~MHz clock are converted to electrical signal and fed to a time-to-digital converter (TDC). The feed-forward signal is also used to trigger a short pulse generator module to generate driven pulse for the phase modulator (PM) for unitary rotation. Electrical polarization controllers (EPC) are used to compensate the polarization drift caused by optical fibre.}
\end{figure}

Fig.~\ref{fig:setup}~b presents details of our experimental realization. Charlie first modulates the continuous wave laser beam ($\lambda = 1552.54$~nm) emitted by a distributed feedback laser (DFB) into 75 ps pulses with an electro-optical modulator (EOM) at a repetition frequency of 100 MHz. After passing the laser pulse through a unbalanced Mach-Zehnder interferometer (MZI) with a path difference of 1 ns, Charlie amplifies the generated two sequential pulses with erbium-doped fibre amplifier (EDFA) and then feeds them into a 300~m dispersion-shifted fibre (DSF) to generate time-bin entangled photon pairs through spontaneous four-wave-mixing (SFWM) process. The DSF is immersed in liquid nitrogen to reduce phonon-related single photon noise. Charlie uses a  filter system composed of cascaded dense wavelength division multiplexing (DWDM) devices to select paired signal (s, 1549.36~nm) and idler (i, 1555.73~nm) photons with the pump light attenuated by 115~dB.  The quantum state of the time-bin entangled photon pairs~\cite{brendel1999pulsed} is $|\Phi^+\rangle_{si}=\frac{1}{\sqrt{2}}(|t_0\rangle_s|t_0\rangle_i+|t_1\rangle_s|t_1\rangle_i)$,where $|t_0\rangle$ and $|t_1\rangle$ represent the first and second time bin, respectively. Charlie then sends the idler photon to Bob and holds the signal photon by propagating it in a 15~km coiled optical fibre. Bob holds the idler photon similarly after he receives it. The temporary storages allow us to perform BSM after entanglement distribution and implement the feed-forward operation in real time.

 Alice generates correlated photon pairs similarly. She obtains single photons by heralding their idler partners. Alice prepares the input quantum state for quantum teleportation, $|\psi\rangle_{in}=\alpha|t_0\rangle_{in}+\beta|t_1\rangle_{in}$, by passing the heralded single photons through an unbalanced MZI with the path difference of 1 ns, and then sends the encoded photon to Charlie.  She also sends the heralding signal (photo-detection signal of idler photons) to Charlie.

To optimize the BSM, the optical delay is carefully arranged between the photon in Charlie's storage and the photon from Alice. Charlie passes the photons through a fibre Bragg grating (FBG) with a bandwidth of 4 GHz before BSM to reduce the spectral distinguishability, which is much smaller than the bandwidth of the pump pulse (8 GHz).  The temporal coherence time of the single photon pulse after FBG is about 110 ps, and the single-photon-state purity is measured to be 0.91(3) and 0.84(2) for photons from Alice and from Charlie, respectively, approximating single eigen mode in the spectral domain. The use of single mode optical fibre ensures the spatial indistinguishability. (FBGs are also used to reduce the bandwidth of idler photons.)

The long coherence time of the prepared quantum state allows us to synchronize independent quantum sources that are far apart with off-the-shelf instruments. To synchronize the independent quantum sources owned by Alice and Charlie, we keep a master clock at the node of Charlie and use it to trigger a pulse patten generator (PPG). The PPG drives the EOM to carve the cw laser beam periodically into 75 ps pulses, separated by 10 ns. These pulses are used to create the EPR pairs for Charlie after amplification. A fraction of each of these optical pulses is sent to Alice's side, where it is detected by a 45 GHz photo-detector. The photo-detector signal is used to trigger the creation of Alice's quantum source.  Using this method, we find that the timing jitter between the laser pulses of Alice and Charlie, quantified by root-mean-square (RMS), is about 2.04 ps, which is much smaller than the 110 ps coherence time of the created single photon pulses. It is evident that this method can be used to synchronize many nodes of a quantum network.

Charlie performs BSM using a fibre beam splitter tree and four superconducting nanowire single photon detectors (SNSPD). The SNSPD has a recovery time of 40 ns which is longer than the difference between time-bins. This detection configuration allows to distinguish all EPR pairs in the state of $|\Psi^-\rangle$ (coupled to $-\mathbf{\sigma}_y|\psi\rangle_{i}$) and 50\% of EPR pairs in the state of $|\Psi^+\rangle$ (coupled to $\mathbf{\sigma}_x|\psi\rangle_{i}$). Synchronized with the master clock at 200 MHz, a field-programmable-gated-array (FPGA) executes the feed-forward logics based on the photo-detection signals from SNSPDs and the heralding signals from Alice. FPGA sends a signal to Bob  for each successful three-fold coincidence detection. At Bob's side, he does not perform unitary rotation on the photon in his storage upon receiving a signal from FPGA if the BSM outcome corresponds to the EPR state $|\Psi^-\rangle$; he performs the unitary rotation on the photon if the signal from FPGA corresponds to the EPR state $|\Psi^+\rangle$ in the BSM. The final state of the photon in Bob's possession is $|\psi\rangle_{fin}=\mathbf{\sigma}_y|\psi\rangle_{i}$. Bob uses an MZI with path difference of 1 ns followed by two SNSPDs to characterize the teleported quantum states~\cite{takesue2015quantum}. The arrival time of the feed-forward signal and the photo-detection signal from the two SNSPDS are recorded with time-to-digital converter (TDC) for coincidence measurement (corresponding to four-fold coincidence detection on the two-photon pairs originally created by Alice and Charlie).

\begin{figure}
\centering
\includegraphics[width=0.6\textwidth]{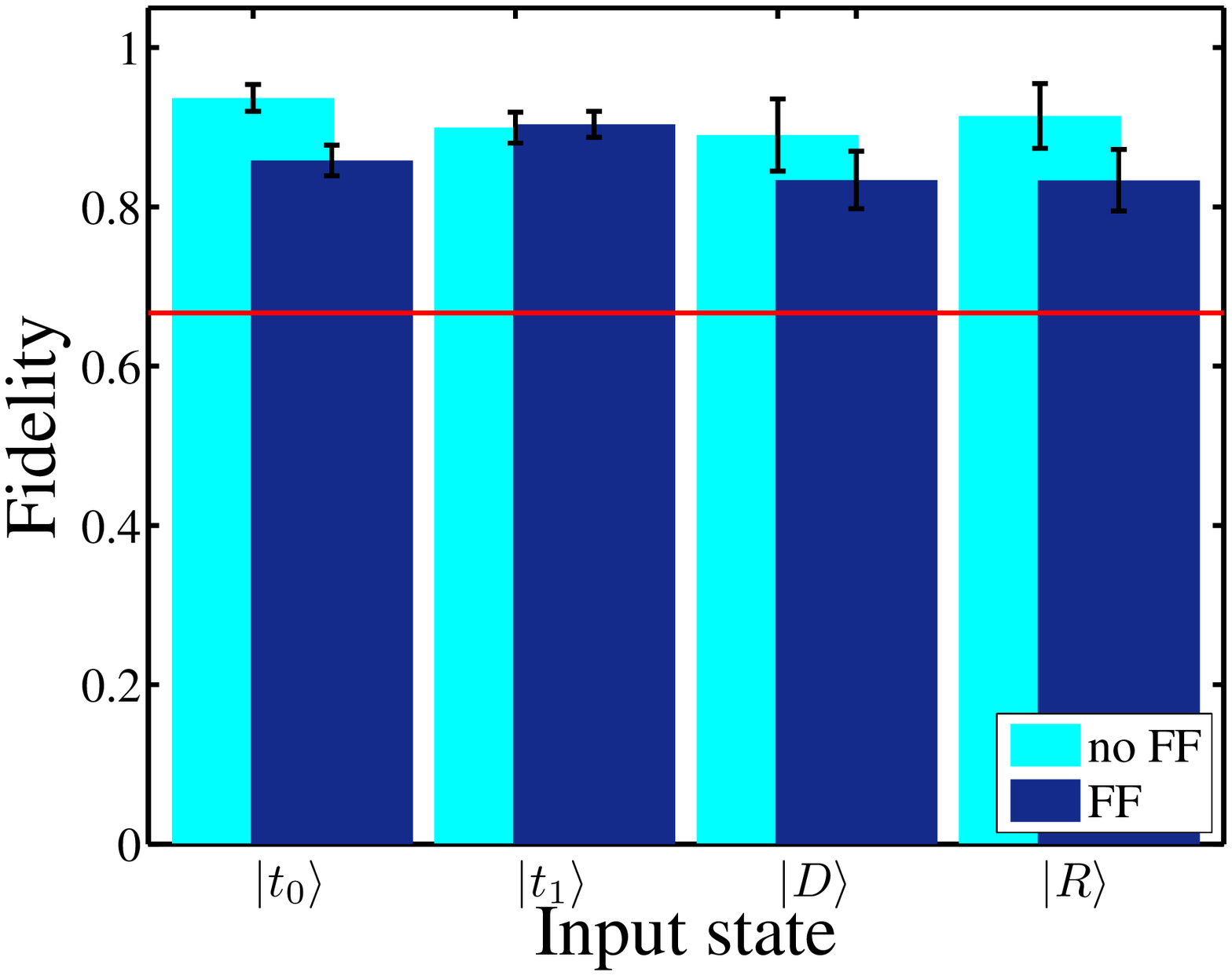}
\caption{\textbf{State fidelities of quantum teleportation with four different input states: $|t_0\rangle$, $|t_1\rangle$, $|D\rangle$ and $|R\rangle$.} The observed state fidelities without and with active feed-forward operation are denoted by color cyan and blue, respectively. All observed state fidelities significantly exceed the classical fidelity limit of 2/3, represented by the red horizontal line.}
\label{fig:fidelitybar}
\end{figure}

\begin{figure}
\centering
\includegraphics[width=0.4\textwidth]{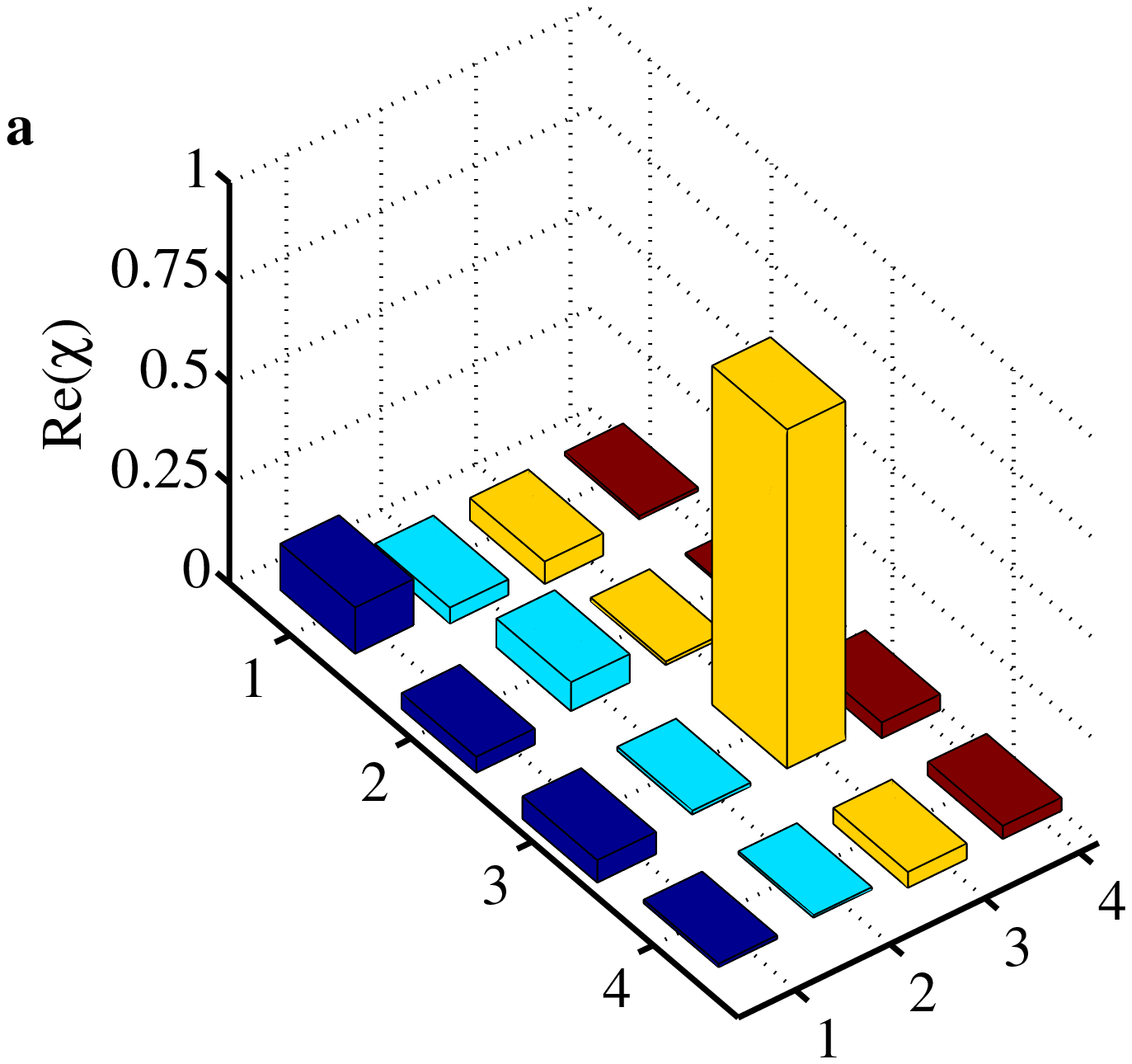}
\includegraphics[width=0.4\textwidth]{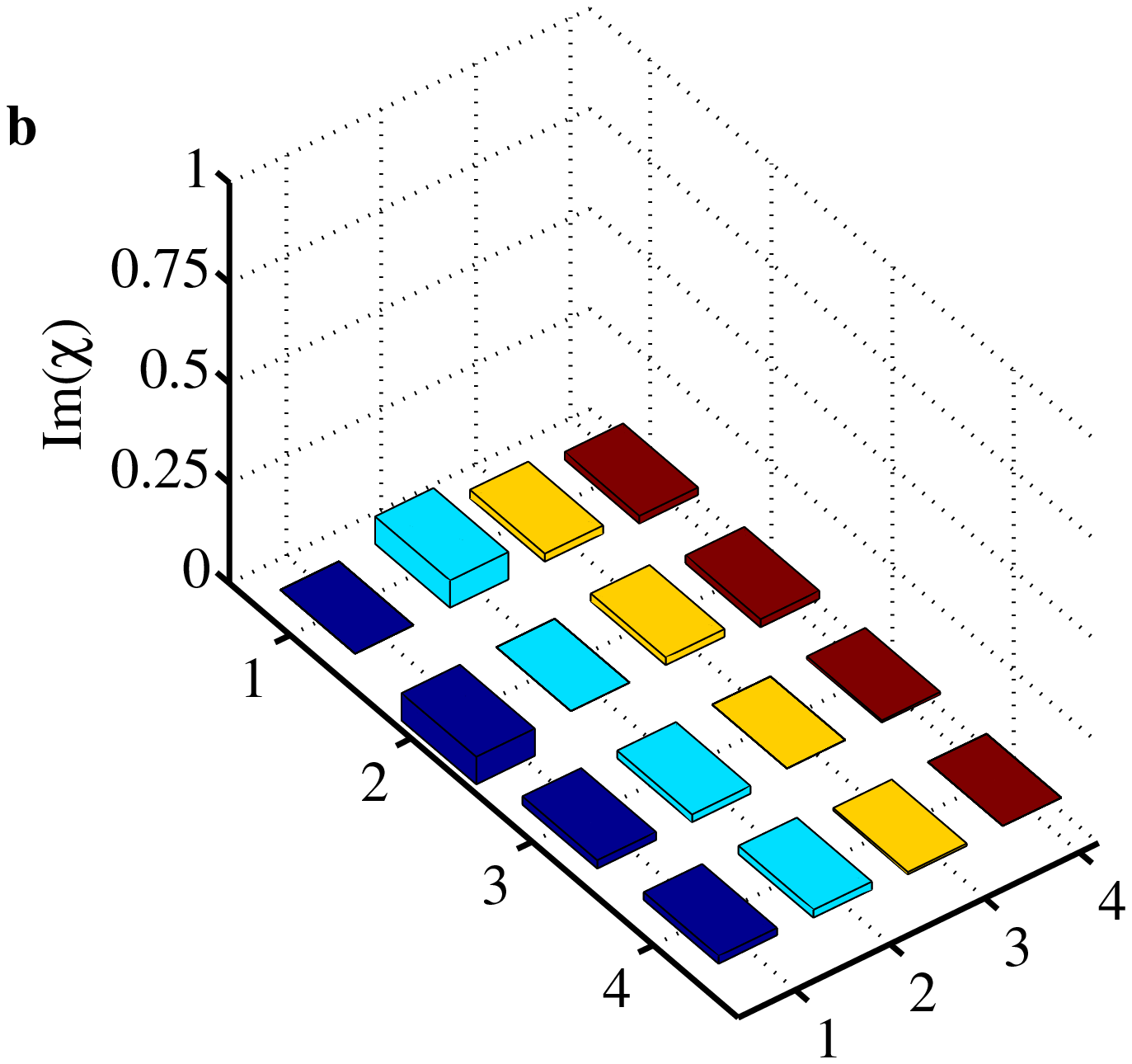}
\\
\includegraphics[width=0.4\textwidth]{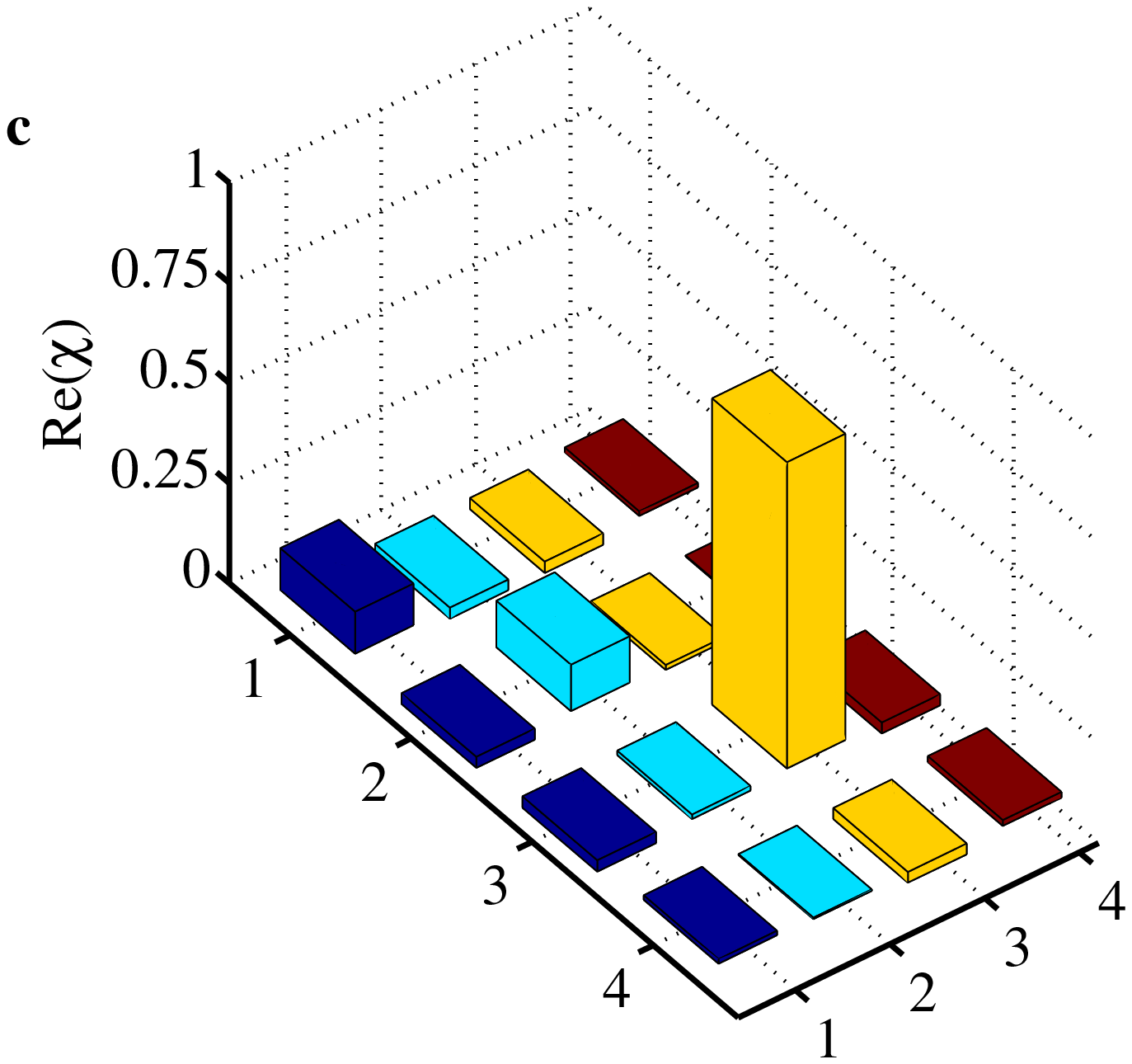}
\includegraphics[width=0.4\textwidth]{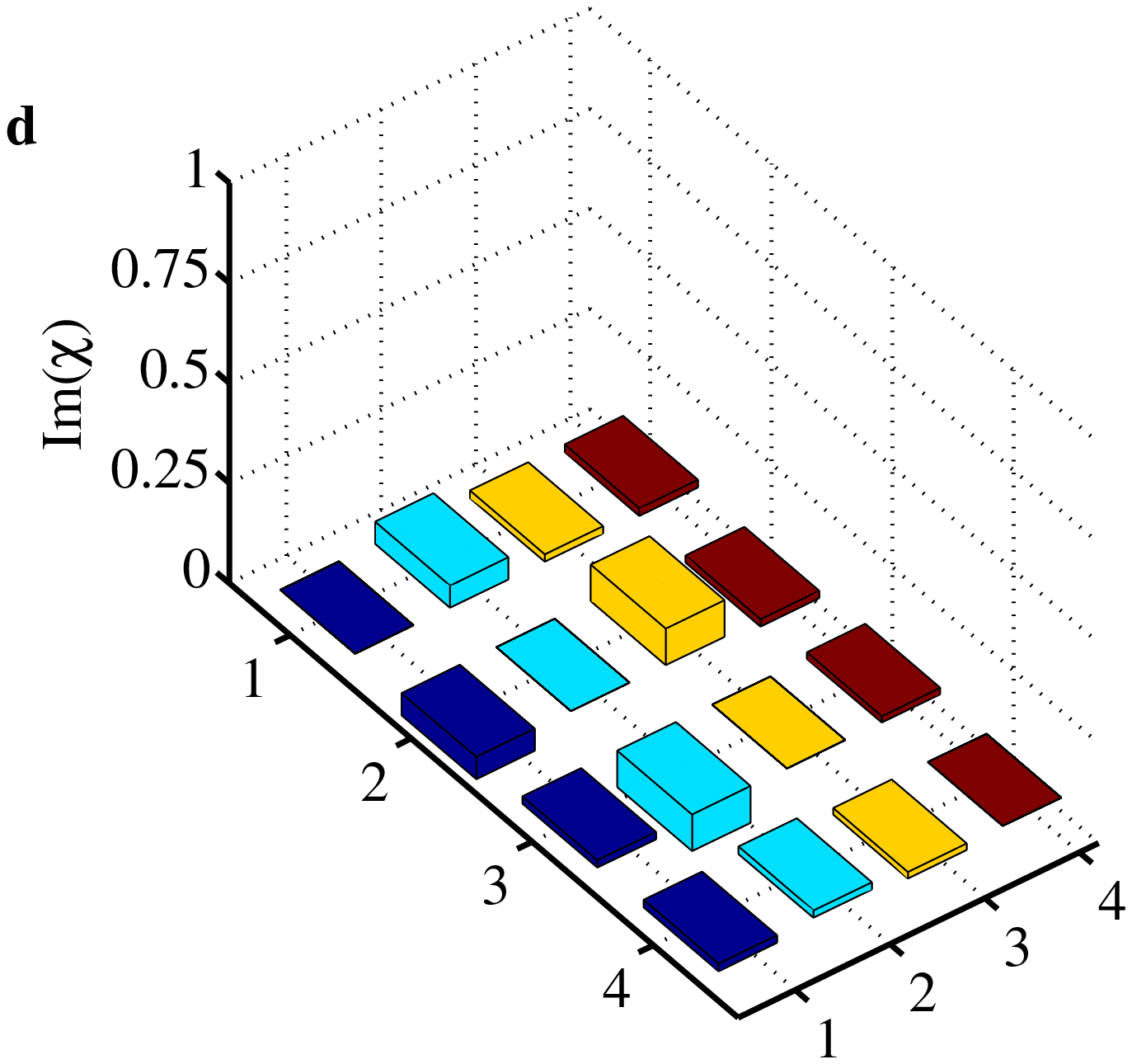}
\caption{\textbf{ Quantum process tomography of quantum teleportation.} \textbf{a}-\textbf{ b} and \textbf{c}-\textbf{d} represent the process matrices for quantum teleportation without and with active feed-forward operation, respectively.}
\label{fig:protom}
\end{figure}

In order to completely characterize the field test of quantum teleportation, we perform quantum state tomography measurement~\cite{takesue2015quantum, james2001measurement} on the teleported quantum states. Without loss of generality, we choose quantum states $|t_0\rangle, |t_1\rangle, |D\rangle$ and $|R\rangle$ as input states~\cite{Ma:143Tele}, where $|D\rangle=\frac{1}{\sqrt{2}}(|t_0\rangle+|t_1\rangle)$ and $|R\rangle=\frac{1}{\sqrt{2}}(|t_0\rangle+i|t_1\rangle)$. The four-fold coincidence rate in our field test is about 2 per hour. We reconstruct density matrices for teleported quantum states, with which we calculate the state fidelities as shown in Fig.~\ref{fig:fidelitybar}. The average quantum state fidelities are $91(3)\%$ and $85(3)\%$ for without feed-forward operation (when the BSM outcome corresponds to  EPR state $|\Psi^-\rangle$ ) and with active feed-forward operation (when the BSM outcome corresponds to either EPR state $|\Psi^-\rangle$ or $|\Psi^+\rangle$), both exceeding the classical limit of 2/3. Following Ref~[\cite{nielsen2010quantum}], we perform quantum process tomography on our experiment and determine the process matrices, which are shown in Fig.~\ref{fig:protom}. The  quantum process fidelity for quantum teleportation in our field test are $84(4)\%$ and $77(3)\%$ for without active feed-forward operation  and with active feed-forward operation. The observed imperfections are partly due to multi-photon process in SFWM and non-ideal unitary rotation. The average photon number per pulse in our experiment is about 0.08 for Alice's quantum source  and 0.03 for Charlie's source, which upper-bounds the visibility of two-photon interference to be 0.917 in our experiment. In addition, the instrumental imperfection restricts the fidelity of unitary rotation to be not better than 0.85.

Lastly, we examine the possibility for any classical process to produce the results observed in out field test. In a classical teleportation, Charlie directly measures the input state and informs Bob to reconstruct the state accordingly, with a maximum state fidelity of 2/3. However, due to a finite number of experimental trials, the statistical fluctuation may allow the classical teleportation to reach and even exceed the state fidelity observed in our experiment. This is quantified by a probability according to Hoeffding's inequality~\cite{hoeffding1963probability},

\begin{equation}
\text{Prob}_\text{classical}(\bar{F}_{\text{classical}}\geq\bar F)\leq \left[\left(
\frac{4/3-2/3}{4/3-\bar{F}}\right)^{\frac{4/3-\bar{F}}{4/3}} \left(\frac{2/3}{\bar{F}} \right)^{\frac{\bar{F}}{4/3}}\right]^{4N},
\label{eq:p-bound}
\end{equation}
where $\bar{F}$ is the state fidelity observed in the experiment and $N$ is the number of experimental trials for each input quantum state. The above inequality reads as that, after $4N$ experimental trials the probability according to any
classical process of predicting an average fidelity $\bar{F}_{\text{classical}}$ no less than the observed average fidelity $\bar{F}$ is no more than $p(N,\bar F)$, which is at most $1.5\times10^{-16}$ with $N=150$ for quantum teleportation without active feed-forward or  $2.4\times10^{-14}$  with $N=240$ for quantum teleportation with active feed-forward in our experiment. With these, we confirm the quantum nature of teleportation over the 30 km optical fibre network.

Our experiment, along with the two recent field tests of quantum teleportation~\cite{Yin:97Tele, Ma:143Tele}, may serve as the benchmark to realize quantum teleportation in the real world. The developed technology is immediately applicable to a wide array of quantum information processing applications with independent quantum sources.

\section*{Acknowledge}
 We are grateful to the staff of the QuantumCTek Co., Ltd. We thank Y.~Liu, H.~Lu, P.~Xu, Y.-P.~Wu, Y.-L.~Tang, X.~Ma, X.~Xie, and M.~Jiang for discussions, and C.~Liu for helping with artwork design. This work was supported by the National Fundamental Research Program (under Grant No. 2011CB921300 and 2013CB336800), the National Natural Science Foundation of China, the Chinese Academy of Science.



\begin{thebibliography}{10}
\expandafter\ifx\csname url\endcsname\relax
  \def\url#1{\texttt{#1}}\fi
\expandafter\ifx\csname urlprefix\endcsname\relax\def\urlprefix{URL }\fi
\providecommand{\bibinfo}[2]{#2}
\providecommand{\eprint}[2][]{\url{#2}}

\bibitem{BB:Teleportation}
\bibinfo{author}{Bennett, C.~H.} \emph{et~al.}
\newblock \bibinfo{title}{Teleporting an unknown quantum state via dual
  classical and einstein-podolsky-rosen channels}.
\newblock \emph{\bibinfo{journal}{Phys. Rev. Lett.}}
  \textbf{\bibinfo{volume}{70}}, \bibinfo{pages}{1895--1899}
  (\bibinfo{year}{1993}).

\bibitem{nielsen2010quantum}
\bibinfo{author}{Nielsen, M.~A.} \& \bibinfo{author}{Chuang, I.~L.}
\newblock \emph{\bibinfo{title}{Quantum computation and quantum information}}
  (\bibinfo{publisher}{Cambridge university press}, \bibinfo{year}{2010}).

\bibitem{Cirac:QuantumNetwork}
\bibinfo{author}{Cirac, J.~I.}, \bibinfo{author}{Zoller, P.},
  \bibinfo{author}{Kimble, H.~J.} \& \bibinfo{author}{Mabuchi, H.}
\newblock \bibinfo{title}{Quantum state transfer and entanglement distribution
  among distant nodes in a quantum network}.
\newblock \emph{\bibinfo{journal}{Phys. Rev. Lett.}}
  \textbf{\bibinfo{volume}{78}}, \bibinfo{pages}{3221--3224}
  (\bibinfo{year}{1997}).

\bibitem{Kimble:QuantumInternet}
\bibinfo{author}{Kimble, H.~J.}
\newblock \bibinfo{title}{The quantum internet}.
\newblock \emph{\bibinfo{journal}{Nature}} \textbf{\bibinfo{volume}{453}},
  \bibinfo{pages}{1023--1030} (\bibinfo{year}{2008}).

\bibitem{barz2012demonstration}
\bibinfo{author}{Barz, S.} \emph{et~al.}
\newblock \bibinfo{title}{Demonstration of blind quantum computing}.
\newblock \emph{\bibinfo{journal}{Science}} \textbf{\bibinfo{volume}{335}},
  \bibinfo{pages}{303--308} (\bibinfo{year}{2012}).

\bibitem{Gisin:Prior}
\bibinfo{author}{Landry, O.}, \bibinfo{author}{van Houwelingen, J. A.~W.},
  \bibinfo{author}{Beveratos, A.}, \bibinfo{author}{Zbinden, H.} \&
  \bibinfo{author}{Gisin, N.}
\newblock \bibinfo{title}{Quantum teleportation over the swisscom
  telecommunication network}.
\newblock \emph{\bibinfo{journal}{. J. Opt. Soc. Am. B}}
  \textbf{\bibinfo{volume}{24}}, \bibinfo{pages}{398--403}
  (\bibinfo{year}{2007}).

\bibitem{Ma:143Tele}
\bibinfo{author}{Ma, X.-S.} \emph{et~al.}
\newblock \bibinfo{title}{Quantum teleportation over 143 kilometres using
  active feed-forward}.
\newblock \emph{\bibinfo{journal}{Nature}} \textbf{\bibinfo{volume}{489}},
  \bibinfo{pages}{269--273} (\bibinfo{year}{2012}).

\bibitem{bouwmeester1997experimental}
\bibinfo{author}{Bouwmeester, D.} \emph{et~al.}
\newblock \bibinfo{title}{Experimental quantum teleportation}.
\newblock \emph{\bibinfo{journal}{Nature}} \textbf{\bibinfo{volume}{390}},
  \bibinfo{pages}{575--579} (\bibinfo{year}{1997}).

\bibitem{boschi1998experimental}
\bibinfo{author}{Boschi, D.}, \bibinfo{author}{Branca, S.},
  \bibinfo{author}{De~Martini, F.}, \bibinfo{author}{Hardy, L.} \&
  \bibinfo{author}{Popescu, S.}
\newblock \bibinfo{title}{Experimental realization of teleporting an unknown
  pure quantum state via dual classical and einstein-podolsky-rosen channels}.
\newblock \emph{\bibinfo{journal}{Phys. Rev. Lett.}}
  \textbf{\bibinfo{volume}{80}}, \bibinfo{pages}{1121} (\bibinfo{year}{1998}).

\bibitem{furusawa1998unconditional}
\bibinfo{author}{Furusawa, A.} \emph{et~al.}
\newblock \bibinfo{title}{Unconditional quantum teleportation}.
\newblock \emph{\bibinfo{journal}{Science}} \textbf{\bibinfo{volume}{282}},
  \bibinfo{pages}{706--709} (\bibinfo{year}{1998}).

\bibitem{nielsen1998complete}
\bibinfo{author}{Nielsen, M.~A.}, \bibinfo{author}{Knill, E.} \&
  \bibinfo{author}{Laflamme, R.}
\newblock \bibinfo{title}{Complete quantum teleportation using nuclear magnetic
  resonance}.
\newblock \emph{\bibinfo{journal}{Nature}} \textbf{\bibinfo{volume}{396}},
  \bibinfo{pages}{52--55} (\bibinfo{year}{1998}).

\bibitem{Marcikic:Tele}
\bibinfo{author}{Marcikic, I.}, \bibinfo{author}{de~Riedmatten, H.},
  \bibinfo{author}{Tittel, W.}, \bibinfo{author}{Zbinden, H.} \&
  \bibinfo{author}{Gisin, N.}
\newblock \bibinfo{title}{Long-distance teleportation of qubits at
  telecommunication wavelengths}.
\newblock \emph{\bibinfo{journal}{Nature}} \textbf{\bibinfo{volume}{421}},
  \bibinfo{pages}{509--513} (\bibinfo{year}{2003}).

\bibitem{barrett2004deterministic}
\bibinfo{author}{Barrett, M.} \emph{et~al.}
\newblock \bibinfo{title}{Deterministic quantum teleportation of atomic
  qubits}.
\newblock \emph{\bibinfo{journal}{Nature}} \textbf{\bibinfo{volume}{429}},
  \bibinfo{pages}{737--739} (\bibinfo{year}{2004}).

\bibitem{riebe2004deterministic}
\bibinfo{author}{Riebe, M.} \emph{et~al.}
\newblock \bibinfo{title}{Deterministic quantum teleportation with atoms}.
\newblock \emph{\bibinfo{journal}{Nature}} \textbf{\bibinfo{volume}{429}},
  \bibinfo{pages}{734--737} (\bibinfo{year}{2004}).

\bibitem{sherson2006quantum}
\bibinfo{author}{Sherson, J.~F.} \emph{et~al.}
\newblock \bibinfo{title}{Quantum teleportation between light and matter}.
\newblock \emph{\bibinfo{journal}{Nature}} \textbf{\bibinfo{volume}{443}},
  \bibinfo{pages}{557--560} (\bibinfo{year}{2006}).

\bibitem{zhang2006experimental}
\bibinfo{author}{Zhang, Q.} \emph{et~al.}
\newblock \bibinfo{title}{Experimental quantum teleportation of a two-qubit
  composite system}.
\newblock \emph{\bibinfo{journal}{Nature Phys.}} \textbf{\bibinfo{volume}{2}},
  \bibinfo{pages}{678--682} (\bibinfo{year}{2006}).

\bibitem{stevenson2013quantum}
\bibinfo{author}{Stevenson, R.} \emph{et~al.}
\newblock \bibinfo{title}{Quantum teleportation of laser-generated photons with
  an entangled-light-emitting diode}.
\newblock \emph{\bibinfo{journal}{Nature Commun.}} \textbf{\bibinfo{volume}{4}}
  (\bibinfo{year}{2013}).

\bibitem{bussieres2014quantum}
\bibinfo{author}{Bussi{\`e}res, F.} \emph{et~al.}
\newblock \bibinfo{title}{Quantum teleportation from a telecom-wavelength
  photon to a solid-state quantum memory}.
\newblock \emph{\bibinfo{journal}{Nature Photon.}}
  \textbf{\bibinfo{volume}{8}}, \bibinfo{pages}{775--778}
  (\bibinfo{year}{2014}).

\bibitem{hanson2014teleportation}
\bibinfo{author}{Pfaff, W.} \emph{et~al.}
\newblock \bibinfo{title}{Unconditional quantum teleportation between distant
  solid-state quantum bits}.
\newblock \emph{\bibinfo{journal}{Science}} \textbf{\bibinfo{volume}{345}},
  \bibinfo{pages}{532--535} (\bibinfo{year}{2014}).

\bibitem{wang2015quantum}
\bibinfo{author}{Wang, X.-L.} \emph{et~al.}
\newblock \bibinfo{title}{Quantum teleportation of multiple degrees of freedom
  of a single photon}.
\newblock \emph{\bibinfo{journal}{Nature}} \textbf{\bibinfo{volume}{518}},
  \bibinfo{pages}{516--519} (\bibinfo{year}{2015}).

\bibitem{takesue2015quantum}
\bibinfo{author}{Takesue, H.} \emph{et~al.}
\newblock \bibinfo{title}{Quantum teleportation over 100 km of fiber using
  highly efficient superconducting nanowire single-photon detectors}.
\newblock \emph{\bibinfo{journal}{Optica}} \textbf{\bibinfo{volume}{2}},
  \bibinfo{pages}{832--835} (\bibinfo{year}{2015}).

\bibitem{ursin2004communications}
\bibinfo{author}{Ursin, R.} \emph{et~al.}
\newblock \bibinfo{title}{Communications: Quantum teleportation across the
  danube}.
\newblock \emph{\bibinfo{journal}{Nature}} \textbf{\bibinfo{volume}{430}},
  \bibinfo{pages}{849--849} (\bibinfo{year}{2004}).

\bibitem{Yin:97Tele}
\bibinfo{author}{Yin, J.} \emph{et~al.}
\newblock \bibinfo{title}{Quantum teleportation and entanglement distribution
  over 100-kilometre free-space channels}.
\newblock \emph{\bibinfo{journal}{Nature}} \textbf{\bibinfo{volume}{488}},
  \bibinfo{pages}{185--188} (\bibinfo{year}{2012}).

\bibitem{Tao:Synch}
\bibinfo{author}{Yang, T.} \emph{et~al.}
\newblock \bibinfo{title}{Experimental synchronization of independent entangled
  photon sources}.
\newblock \emph{\bibinfo{journal}{Phys. Rev. Lett.}}
  \textbf{\bibinfo{volume}{96}}, \bibinfo{pages}{110501}
  (\bibinfo{year}{2006}).

\bibitem{Rainer:Synch}
\bibinfo{author}{Kaltenbaek, R.}, \bibinfo{author}{Blauensteiner, B.},
  \bibinfo{author}{\ifmmode~\dot{Z}\else \.{Z}\fi{}ukowski, M.},
  \bibinfo{author}{Aspelmeyer, M.} \& \bibinfo{author}{Zeilinger, A.}
\newblock \bibinfo{title}{Experimental interference of independent photons}.
\newblock \emph{\bibinfo{journal}{Phys. Rev. Lett.}}
  \textbf{\bibinfo{volume}{96}}, \bibinfo{pages}{240502}
  (\bibinfo{year}{2006}).

\bibitem{Halder:TimeMeasurement}
\bibinfo{author}{Halder, M.} \emph{et~al.}
\newblock \bibinfo{title}{Entangling independent photons by time measurement}.
\newblock \emph{\bibinfo{journal}{Nature Phys.}} \textbf{\bibinfo{volume}{3}},
  \bibinfo{pages}{692--695} (\bibinfo{year}{2007}).

\bibitem{patel2010two}
\bibinfo{author}{Patel, R.~B.} \emph{et~al.}
\newblock \bibinfo{title}{Two-photon interference of the emission from
  electrically tunable remote quantum dots}.
\newblock \emph{\bibinfo{journal}{Nature Photon.}}
  \textbf{\bibinfo{volume}{4}}, \bibinfo{pages}{632--635}
  (\bibinfo{year}{2010}).

\bibitem{brendel1999pulsed}
\bibinfo{author}{Brendel, J.}, \bibinfo{author}{Gisin, N.},
  \bibinfo{author}{Tittel, W.} \& \bibinfo{author}{Zbinden, H.}
\newblock \bibinfo{title}{Pulsed energy-time entangled twin-photon source for
  quantum communication}.
\newblock \emph{\bibinfo{journal}{Phys. Rev. Lett.}}
  \textbf{\bibinfo{volume}{82}}, \bibinfo{pages}{2594} (\bibinfo{year}{1999}).

\bibitem{james2001measurement}
\bibinfo{author}{James, D.~F.}, \bibinfo{author}{Kwiat, P.~G.},
  \bibinfo{author}{Munro, W.~J.} \& \bibinfo{author}{White, A.~G.}
\newblock \bibinfo{title}{Measurement of qubits}.
\newblock \emph{\bibinfo{journal}{Phys. Rev. A}} \textbf{\bibinfo{volume}{64}},
  \bibinfo{pages}{052312} (\bibinfo{year}{2001}).

\bibitem{hoeffding1963probability}
\bibinfo{author}{Hoeffding, W.}
\newblock \bibinfo{title}{Probability inequalities for sums of bounded random
  variables}.
\newblock \emph{\bibinfo{journal}{J. Amer. Statist. Assoc.}}
  \textbf{\bibinfo{volume}{58}}, \bibinfo{pages}{13--30}
  (\bibinfo{year}{1963}).

\end{thebibliography}
\end{document}